\documentclass{snmp2003}
\renewcommand{\Part}{Part~2}
\renewcommand{\Volume}{Vol.~50}
\begin{document}

\FirstPageHeading{Kotel'nikov}

\ShortArticleName{On the Possibility of Faster-Than-Light Motions}

\ArticleName{On  the  Possibility   of  Faster-Than-Light  Motions\\
in Nonlinear Electrodynamics}

\Author{Gennadii KOTEL'NIKOV}

\AuthorNameForHeading{G.   Kotel'nikov}
\AuthorNameForContents{KOTEL'NIKOV G.}
\ArticleNameForContents{On  the  Possibility   of  Faster-Than-Light  Motions
in Nonlinear\\ Electrodynamics}

\Address{RRC Kurchatov  Institute,  1 Kurchatov  Sq.,  123182 Moscow,
Russia}

 \Email{kga@electronics.kiae.ru}

\Abstract{A version   of   electrodynamics  is  constructed  in  which
faster-than-light motions of electromagnetic fields and particles with
real masses are possible.}

\section{Introduction}  For  certainty,  by
faster-than-light  motions  we  shall  understand  the  motions   with
velocities $v>3\cdot10^{8}$~m/sec.  The existence of such motions is
the question discussed in modern physics.

As already  as in      1946--1948 Blokhintsev
\cite{kotel'nikov:blokhintsev46}  paid  attention  to  the  possibility  of
formulating the   field   theory   that allows  propagation    of
faster-than-light  (superluminal) interactions outside the light cone.
Some time later  he  also  noted  the  possibility  of  the  existence  of
superluminal  solutions  in  nonlinear  equations  of  electrodynamics
\cite{kotel'nikov:blokhintsev52}.  Kirzhnits \cite{kotel'nikov:kirzhnits} showed
that     a     particle     possessing     the    tensor    of    mass
${M^i}_k={\rm diag}\,(m_0,m_1,m_1,m_1)$, $i,k=0,1,2,3$, $g_{ab}={\rm diag}\,(+,-,-,-)$
can move      faster-than-light,      if      $m_0>m_1$.     Terletsky
\cite{kotel'nikov:terletsky}  introduced  into  theoretical   physics   the
particles with imaging rest masses moving faster-than-light.  Feinberg
\cite{kotel'nikov:feinberg} named these particles  tachyons  and  described
their main properties.

Research on  superluminal  tachyon  motions   opened   up   additional
opportunities  which  were  studied  by  many authors,  for example by
Bilaniuk  and   Sudarshan   \cite{kotel'nikov:bilaniuk&sudarshan},   Recami
\cite{kotel'nikov:recami},  Mignani (see  \cite{kotel'nikov:recami}), Kirzhnits
and    Sazonov    \cite{kotel'nikov:kirzhnits&sazonov},    Corben     (see
\cite{kotel'nikov:recami}),      Patty     \cite{kotel'nikov:patty},     Oleinik
\cite{kotel'nikov:oleinik}.  It has led to  original  scientific  direction
(several hundred publications).  The tachyon movements may formally be
described by Special Relativity (SR) expanded to the domain of motions
$s^2<0$. For comparison,  the standard theory describes motions on the
light cone   $s^2=0$   and   in   the   domain   $s^2>0$  when  $v\leq
3\cdot10^{8}$~m/sec.

The publications are also known in which the violation  of  invariance
of   the   speed   of   light   is  considered
\cite{kotel'nikov:pauli,kotel'nikov:logunov,
kotel'nikov:glashow,kotel'nikov:rapier,kotel'nikov:hsu,
kotel'nikov:chubykalo&smirnov-rueda,kotel'nikov:kotel'nikov}.  One   can   note,   for   example,   Pauli
monograph~\cite{kotel'nikov:pauli}  where  elements  of  Ritz  and  Abraham
theory are presented;  Logunov's lectures \cite{kotel'nikov:logunov} in which
SR formulation     in     affine     space was given;    Glashow's
paper~\cite{kotel'nikov:glashow} discussing  experimental  consequences  of
the  violation  of  Lorentz-invariance  in astrophysics;  publications
\cite{kotel'nikov:rapier,kotel'nikov:hsu,
kotel'nikov:chubykalo&smirnov-rueda,kotel'nikov:kotel'nikov}  on  the  violation  of
invariance of the speed of light in SR.

A version  of the theory permitting faster-than-light motions
of electromagnetic fields and charged particles with  real  masses  is
proposed  below as continuation of such investigations.

\section{Formal construction of the theory}

Let us introduce the space-time ${\mathbb R}^4$,  metric properties of which may
depend on the velocity of a~particle being investigated and  take  the
metric of ${\mathbb R}^4$ in the form:
\begin{gather}
ds^2=\big({c_0}^2+v^2\big)dt^2-dx^2-dy^2-dz^2                 \nonumber\\
\phantom{ds^2}{}=\big({{c'}_0}^2+{v'}^2\big)(dt')^2-(dx')^2-(dy')^2-(dz')^2 -
{\rm invariant}.\label{kotel'nikov:eq1}
\end{gather}
Here $x$,  $y$,  $z$ are the spatial coordinates, $t$ is the time, $c_0$,
${c'}_0$ are the proper values of the speed of light, $v$ is the velocity
of a body under study with respect to the reference frame $K$.  Let us
connect the co-moving frame $K'$ with this body.  Let the proper speed
of light be invariant
\begin{equation}\label{kotel'nikov:eq2}
c_0={c'}_0=3\cdot10^{8}~{\rm m/sec} - {\rm invariant}.
\end{equation}
As a result,  the common time similar to Newton one may be  introduced
on the  trajectory  of the frame $K'$ (when ${\boldsymbol v}=d{\boldsymbol x}/dt$) and
the velocity of light $c$, corresponding to the velocity ${\boldsymbol v}$:
\begin{equation}\label{kotel'nikov:eq3}
dt={dt'}_0 \to t={t'}_0,
\end{equation}
\begin{equation}\label{kotel'nikov:eq4}
c=\pm c_0\sqrt{1+\frac{v^2}{{c_0}^2}}.
\end{equation}
We shall name the value $c_0=3\cdot10^{8}$ m/sec the speed of light,
the   value  $c$ -- the velocity of light\footnote{In  the  form  of  $c'=c(1-\beta^2)^{1/2}$
expression (\ref{kotel'nikov:eq4}) was  obtained  by  Abraham  in  the
model  of  Aether \cite{kotel'nikov:pauli}.}.  In
accordance  with  the  hypothesis  of  homogeneity  and  isotropy   of
space-time, the velocity $v$ of a free particle does not depend on $t$
and ${\boldsymbol x}$.  The velocity of  light  (\ref{kotel'nikov:eq4})  is  a
constant in this case. Let $dx^0=cdt$ and
\begin{equation}\label{kotel'nikov:eq5}
x^0=\int_{0}^{t}cd\tau=\pm\int_{0}^{t}
c_0\sqrt{1+\frac{v^2}{{c_0}^2}} d\tau
\end{equation}
be the ``time'' $x^0$ when $v\ne {\rm const}$ also.  Keeping this in  mind,  we
rewrite the expression (\ref{kotel'nikov:eq1}) in the form
\begin{equation}\label{kotel'nikov:eq6}
ds^2=\big(dx^0\big)^2-\big(dx^1\big)^2-\big(dx^2\big)^2-\big(dx^3\big)^2,
\end{equation}
where $x^{1,2,3}=(x,y,z)$.     As     is     known,     the     metric
(\ref{kotel'nikov:eq6})   describes  the  flat  homogeneous  Minkowski
space-time $M^4$ with $g_{ik}={\rm diag}\,(+,-,-,-)$,  $i,k=0,1,2,3$.  Under
the   condition   (\ref{kotel'nikov:eq3}),  infinitesimal   space-time
transformations,      retaining     invariance     of     the     form
(\ref{kotel'nikov:eq6}),  are accompanied by the transformation of the
velocity of light \cite{kotel'nikov:kotel'nikov}:
\begin{equation}\label{kotel'nikov:eq7}
{dx'}^i={L^i}_k dx^k, \qquad c'=c(1-{\boldsymbol \beta}\cdot{\boldsymbol u})/\sqrt{1-\beta^2},
\qquad i,k = 0,1,2,3.
\end{equation}
Here ${L^i}_k$  is  the matrix of Lorentz group with ${\boldsymbol \beta}={\boldsymbol
V}/c={\bf   const}$   \cite{kotel'nikov:pauli},   ${\boldsymbol   u}={\boldsymbol   v}/c$.
Corresponding  homogeneous integral transformations in the case of the
one-parametric Lorentz group $L_1$ are
\begin{equation}\label{kotel'nikov:eq11}
{x'}^0=\frac{x^0-\beta     x^1}{\sqrt{1-\beta^2}},\qquad
{x'}^1=\frac{x^1-\beta   x^0}{\sqrt{1-\beta^2}},\qquad   {x'}^2=x^2,\qquad
{x'}^3=x^3,  \qquad  c'=c\frac{1-\beta u^1}{\sqrt{1-\beta^2}},
\end{equation}
with ${\boldsymbol   \beta}=(V/c,0,0)$,  $u^1=dx^1/dx^0=v_x/c$.  For  inertial
motions in the space-time (\ref{kotel'nikov:eq1}) we have
\[
t'=t,\qquad  x'=\frac{x-Vt}{\sqrt{1-V^2/c^2}},\qquad  y'=y,\qquad   z'=z,\qquad
c'=c\frac{1-Vv_x/c^2}{\sqrt{1-V^2/c^2}},
\]
where we   take    into   account   $v_x=x/t$.   The   transformations
(\ref{kotel'nikov:eq11})    are    induced     by     the     operator
$X=x_1\partial_0-x_0\partial_1-u^1c\partial_c$  which  is  the  sum of
Lorentz group $L_1$ generator $J_{01}=x_1\partial_0-x_0\partial_1$ and
the   generator   $D=c\partial_c$   of   scale  transformations  group
$\triangle_1$ of the velocity of light $c'=\gamma c$.  We can say that
these generators act in 5-space $M^4\times V^1$ where $V^1$ is a subspace of
the   velocities   of   light,   and    that    the    transformations
(\ref{kotel'nikov:eq11})   belong  to  the  group  of  direct  product
$L_1\times \triangle_1$. The generators $J_{01}$ and $D$ and transformations
(\ref{kotel'nikov:eq11})  are  respectively the symmetry operators and
symmetry transformations for the equation of the zero cone $s^2=0$  in
the  5-space  $V^5=M^4\times V^1$  where  $|c|<\infty$  includes  the subset
$c_0<|c|<\infty$:
\[
s^2=\big(x^0\big)^2-\big(x^1\big)^2-\big(x^2\big)^2-\big(x^3\big)^2=0,\qquad
J_{01}\big(s^2\big)=0, \quad D\big(s^2\big)=0,  \quad  [J_{01},D]=0.
\]
The relationships  between  the  partial  derivatives of the variables
$(t,x,y,z)$ and $(x^0,x^1,x^2,x^3)$ are as follows:
\begin{gather}
\frac{\partial}{\partial t}=\frac{\partial               x^0}{\partial
t}\frac{\partial}{\partial      x^0}+      \sum_{\alpha}\frac{\partial
x^{\alpha}}{\partial            t}\frac{\partial}            {\partial
x^{\alpha}}=c\frac{\partial}{\partial x^0},                       \nonumber\\
\frac{\partial}{\partial x}=\frac{\partial      x^0}{\partial       x}
\frac{\partial}{\partial  x^0}+\sum_{\alpha}\frac{\partial x^{\alpha}}
{\partial           x}\frac{\partial}{\partial            x^{\alpha}}=
\left(\int\frac{\partial    c}{\partial   x}d\tau\right)   \frac{\partial
}{\partial x^0}+ \frac{\partial}{\partial x^1}, \qquad \alpha=1,2,3.\label{kotel'nikov:eq13}
\end{gather}
The expressions  for  $\partial/\partial  y$ and $\partial/\partial z$
are analogous to the  expression  $\partial/\partial  x$.  Further  we
restrict ourselves by the case of positive values of the velocities of
light and by studying a variant of the theory in which the velocity of
light  in the range of interactions may only depend on the time~``$t$'',
i.e.\ $c=c(t)\leftrightarrow c=c(x^0)$.  The relationship between $x^0$
and    $t$    may    be    deduced    from   the   solution   of  equation
(\ref{kotel'nikov:eq5}). Then
\begin{equation}\label{kotel'nikov:eq14}
\displaystyle
\nabla c\big(x^0\big)=0, \quad  c=c\big(x^0\big), \quad u^2=u^2\big(x^0\big) \quad
\leftrightarrow \quad \nabla c(t)=0, \quad c=c(t),  \quad v^2=v^2(t).
\end{equation}

Let us note some features of motions in this case.

1.~As in  SR,  the parameter $\beta=V/c$ in the present work is in the
range $0\leq\beta<1$.

2.~As in SR, the value $dx^0$ is the exact differential in view of the
condition $\nabla c=0$.

3.~As distinct  from SR,  the ``time'' $x^0=ct$ in the present work is a
function of the time $t$ only for the case of a free particle.  In the
range of interaction the velocity of light may depend on time $t$, and
the value $x^0$ becomes the functional (\ref{kotel'nikov:eq5}) of  the
function $c(t)$.

4.~The parameter      $\beta=V/c$      of      the     transformations
(\ref{kotel'nikov:eq11}) may be constant  not  only  at  the  constant
velocity of light,  but also with $c=c(t)$.  Indeed we may accept that
$0\leq\beta=V(t)/c_0(1+v^2(t)/{c_0}^2)^{1/2}={\rm const}<1$, which is not
in contradiction with $V=V(t), \ c=c(t)$. This property permits one to
use the matrix ${L^i}_k$ from (\ref{kotel'nikov:eq7}) for constructing
Lorentz invariants in the range of interaction where $c=c(t)$.

5.~The condition  (\ref{kotel'nikov:eq14})   is   invariant   on   the
trajectory  of  a  particle  because  $t'=t$  in this case.  Replacing
space-time variables in the expression $\nabla c'=0$  ($c'=\gamma  c$)
we  find that the condition $\nabla c=0$ follows from $\nabla c'=0$ if
$\nabla\gamma=0  \to   (\beta-u^1)\partial\beta/\partial   x^{\alpha}+
(\beta^3-\beta)\partial  u^1/\partial  x^{\alpha}=0$,  $\alpha=1,2,3$.
The system contains  the  solution  $\beta={\rm const}$,  $u^1=u^1(x^0)$  in
agreement with (\ref{kotel'nikov:eq14}) and item 4.

Keeping this   in   mind,   let   us   construct    in    the    space
(\ref{kotel'nikov:eq6}) a theory like SR, reflect it on the space-time
(\ref{kotel'nikov:eq1})     by     means     of      the      formulas
(\ref{kotel'nikov:eq13}), (\ref{kotel'nikov:eq14})  and  consider  the
main properties of the theory.
Following \cite{kotel'nikov:landau&lifshitz}, we may construct the integral
of action in the form:
\begin{gather}
S={S}_m+{S}_{mf}+{S}_f=
-mc_0\int ds - \frac{e}{c_0}\int A_i dx^i-
\frac{1}{16\pi c_0}\int F_{ik}F^{ik} d^4 x    \nonumber\\
\phantom{S}=\int\left[-mc_0\sqrt{1-u^2}+\frac{e}{c_0}
({\boldsymbol A}\cdot{\boldsymbol u}-\phi)\right]dx^0-
\frac{1}{8\pi c_0}\int(E^2-H^2)d^3xdx^0\nonumber                           \\
\phantom{S}=-mc_0\int ds - \frac{1}{c_0}\int A_ij^id^4x-
\frac{1}{16\pi c_0}\int F_{ik}F^{ik} d^4x.\label{kotel'nikov:eq15}
\end{gather}
Here in accordance with  \cite{kotel'nikov:landau&lifshitz}  $S_m=-mc_0\int
ds=-mc_0\int(c_0/c)dx^0=-mc_0\int(1-u^2)^{1/2}dx^0$  is the action for
a free particle;  $S_f=-(1/16\pi  c_0)\int  F_{ik}F^{ik}d^4x$  is  the
action    for   free   electromagnetic   field,   $S_{mf}=-(e/c_0)\int
A_idx^i=-(1/c_0)\int  A_ij^id^4x$  is  the  action  corresponding   to
the interaction   between   the   charge   $e$   of   a  particle  and
electromagnetic field;  $A^i=(\phi,{\boldsymbol A})$  is  the  4-potential;
$A_i=g_{ik}A^k$;  $j^i=(\rho,\rho{\boldsymbol  u})$ is the 4-vector of current
density;  $\rho$ is the charge density;  $F_{ik}=\partial A_k/\partial
x^i-\partial A_i/\partial x^k$ is the tensor of electromagnetic field;
$i,k=0,1,2,3$;  ${\boldsymbol E}=-\partial {\boldsymbol A}/\partial x^0-\nabla\phi$ is
the  electrical field;  ${\boldsymbol H}=\nabla\times {\boldsymbol A}$ is the magnetic
field;  $F_{ik}F^{ik}=2(H^2-E^2)$;  $dx^4=dx^0dx^1dx^2dx^3$   is   the
element of the invariant 4-volume.  The speed of light $c_0$, the mass
$m$, the electric charge $e$ are invariant constants of the theory.

In spite   of  the  similarity,  the  action  (\ref{kotel'nikov:eq15})
differs from  the  action  of  SR  \cite{kotel'nikov:landau&lifshitz}.  The
current   density   was   taken   in   the   form  $j^i=(\rho,\rho{\boldsymbol
u})=(\rho,\rho  {\boldsymbol v}/c)$  instead  of   $j^i=(\rho,\rho{\boldsymbol   v})$
\cite{kotel'nikov:landau&lifshitz}.  The electromagnetic field was taken in
the      form      ${\boldsymbol      E}=-\partial      {\boldsymbol      A}/\partial
x^0-\nabla\phi=-(1/c)\partial  {\boldsymbol  A}/\partial t-\nabla\phi$ instead
of   ${\boldsymbol   E}=-(1/c_0)\partial   {\boldsymbol   A}/\partial    t-\nabla\phi$
\cite{kotel'nikov:landau&lifshitz}.      The     current     density     in
(\ref{kotel'nikov:eq15}) is similar to the one  from  Pauli  monograph
\cite{kotel'nikov:pauli}  with  the  only  difference  that  the  3-current
density in (\ref{kotel'nikov:eq15})  is  $\rho{\boldsymbol  v}/c$  instead  of
being $\rho{\boldsymbol v}/c_0$ \cite{kotel'nikov:pauli}. Analogously, the velocity
of 4-potential propagation in  (\ref{kotel'nikov:eq15})  is  $c$  from
(\ref{kotel'nikov:eq4})       instead      of       $c_0$           in
\cite{kotel'nikov:landau&lifshitz}.

In addition to Lorentz-invariance  \cite{kotel'nikov:landau&lifshitz},  the
action  (\ref{kotel'nikov:eq15}) is also invariant with respect to any
transformations of the  velocity  of  light  and,  consequently,  with
respect to the transformations $c'=\gamma c$,  as the value $c$ is not
contained in the expression (\ref{kotel'nikov:eq15}).  As a result the
action  (\ref{kotel'nikov:eq15})  is  invariant  with  respect  to the
transformations (\ref{kotel'nikov:eq11})  from  the  group  of  direct
product   $L_1\times \triangle_1\subset   L_6\times \triangle_1$,  containing  the
Lorentz group  $L_6$ and the scale group $\triangle_1$ ($c'=\gamma c$)
as subgroups.

Lagrangian $L$, generalized 4-momentum ${\boldsymbol P}$ and
generalized energy  ${\cal H}$ of a particle are:
\begin{gather}\label{kotel'nikov:eq16}
L=-mc_0\sqrt{1-u^2}+\frac{e}{c_0}({\boldsymbol A}\cdot{\boldsymbol u}-\phi),
\\
\label{kotel'nikov:eq17}
{\boldsymbol P}=\frac{\partial  L}{\partial{\boldsymbol u}}=\frac{mc_0{\boldsymbol u}}{\sqrt
{1-u^2}}+\frac{e}{c_0}{\boldsymbol A}={\boldsymbol p}+\frac{e}{c_0}{\boldsymbol A},
\\
\label{kotel'nikov:eq18}
{\cal H}={\boldsymbol P}\cdot{\boldsymbol u}-L=\frac{mc_0}{\sqrt{1-u^2}}+
\frac{e\phi}{c_0}=\frac{{\cal E}}{c_0}+\frac{e\phi}{c_0}.
\end{gather}
Here ${\boldsymbol p}=m{\boldsymbol v}$ is  the  momentum,  ${\cal  E}=mc_0c$  is  the
energy,  ${\cal E}_0=m{c_0}^2$ is the rest energy of a~particle. As in
SR,  the values ${\cal E}$ and  ${\boldsymbol  p}$  may  be  united  into  the
4-momentum $p^i$:
\begin{equation}\label{kotel'nikov:eq19}
p^i=mc_0u^i=\left(\frac{mc_0}{\sqrt{1-u^2}},\frac{mc_0u^{\alpha}}
{\sqrt{1-u^2}}\right)=\left(\frac{{\cal E}}{c_0},m{\boldsymbol v}\right), \qquad
\alpha=1,2,3.
\end{equation}
The components of $p^i$ are related by the expressions:
\begin{gather}
p_ip^i=\frac{{\cal E}^2}{{c_0}^2}-{\boldsymbol p}^2=m^2{c_0}^2, \nonumber\\
{\boldsymbol p}=\frac{{\cal E}}{c_0c}{\boldsymbol v}\qquad \mbox{or}\qquad
{\boldsymbol p}=\frac{{\cal E}}{c_0}{\boldsymbol n},\quad
 {\boldsymbol n}=\frac{{\boldsymbol c}}{c}, \quad \mbox{if}\quad
 m=0, \quad {\boldsymbol v}={\boldsymbol c}.\label{kotel'nikov:eq20}
\end{gather}
One can see from here that the momentum of a particle  with  the  zero
mass $m=0$  is  independent  of  the  particle velocity $v=c$ and only
determined by the particle energy.  As in SR,  for  the  case  of  the
photon we find: $p^i=({\hbar\omega}/{c_0},{\hbar\omega}{\boldsymbol n}/{c_0})=
({\cal E}/{c_0},\hbar{\boldsymbol k})$  where  ${\cal  E}=\hbar\omega$,  ${\boldsymbol
k}=(\omega/c_0){\boldsymbol n}$,  $\omega$ is the frequency of electromagnetic
field, $\hbar$ is the Planck constant.

For constructing the equations of motion for a  charged  particle  and
electromagnetic    field    let   us   start   from   the   mechanical
\cite{kotel'nikov:landau&lifshitz}  and  the   field   Lagrange   equations
\cite{kotel'nikov:ivanenko&sokolov, kotel'nikov:bogoliubov&shirkov}
\begin{equation}\label{kotel'niov:eq21}
\frac{d}{dx^0}\frac{\partial L}{\partial{\boldsymbol u}}-
\frac{\partial L}{\partial{\boldsymbol x}}=0,\qquad
\displaystyle
\frac{\partial}{\partial x^k}\frac{\partial{\cal L}}
{\partial(\partial
A_i/\partial   x^k)}-
\frac{\partial{\cal  L}}{\partial  A_i}=0.
\end{equation}
Here $L$    is     Lagrangian     (\ref{kotel'nikov:eq16}),     ${\cal
L}=-(1/c_0)A_ij^i-(1/16\pi c_0) F_{ik}F^{ik}$. Taking into account the
equality  $\nabla({\boldsymbol  a}  \cdot{\boldsymbol   b})
=({\boldsymbol   a}\cdot\nabla){\boldsymbol
b}+({\boldsymbol  b}\cdot\nabla){\boldsymbol  a}+  {\boldsymbol  a}
\times (\nabla\times {\boldsymbol
b})+{\boldsymbol b}\times (\nabla\times {\boldsymbol a})$, the permutational ratios of
the     tensor    of    electromagnetic    field,    the    expression
$\partial(F^{ik}F_{ik})/\partial(\partial A_i/\partial  x^k)=-4F^{ik}$
\cite{kotel'nikov:landau&lifshitz}        we       obtain       $d(mc_0{\boldsymbol
u}/(1-u^2)^{1/2})/dx^0=(e/c_0){\boldsymbol E}+(e/c_0){\boldsymbol u}\times {\boldsymbol  H}$,
$\partial  F_{ik}/\partial  x^l+\partial  F_{kl}/\partial x^i+\partial
F_{li}/\partial  x^k=0$,  $\partial  F^{ik}/\partial  x^k+4\pi  j^i=0$
\cite{kotel'nikov:landau&lifshitz}.      With     help     of     relations
(\ref{kotel'nikov:eq13}),  (\ref{kotel'nikov:eq14})  and   expressions
$dx^0=cdt,   \   (1-u^2)^{1/2}=c_0/c$  we  may  obtain  the  following
equations  of  motions  in  the   space-time   (\ref{kotel'nikov:eq1})
\cite{kotel'nikov:kotel'nikov}:
\begin{gather}
\frac{d{\boldsymbol p}}{dt}=m\frac{d{\boldsymbol v}}{dt}=\frac{c}{c_0}e{\boldsymbol E}+
\frac{e}{c_0}{\boldsymbol v}\times {\boldsymbol H},\nonumber                                \\
\frac{d{\cal E}}{dt}=e{\boldsymbol E}\cdot{\boldsymbol v}\to m\frac{dc}{dt}=
\frac{e}{c_0}{\boldsymbol v}\cdot{\boldsymbol E};\label{kotel'nikov:eq22}
\\
\nabla \times {\boldsymbol E}+\frac{1}{c}\frac{\partial{\boldsymbol H}}{\partial t}=0,\qquad
\displaystyle
\nabla\cdot{\boldsymbol H}=0,\nonumber                                                \\
\nabla \times {\boldsymbol H}-\frac{1}{c}\frac{\partial{\boldsymbol E}}{\partial t}=
\displaystyle
4\pi\rho \frac{{\boldsymbol v}}{c},
\qquad  \nabla\cdot{\boldsymbol E}=4\pi\rho.\label{kotel'nikov:eq24}
\end{gather}
Equations (\ref{kotel'nikov:eq22})  determine  dynamics of motion of a
charged  particle  with  a  mass  $m$  in  external  field.  The   set
(\ref{kotel'nikov:eq24})   consists   of  Maxwell  equations  for  the
velocity  of  light   $c=c_0(1+v^2/{c_0}^2)^{1/2}$   with   $v={\rm const}$.
Considered  together,  they  form  the  set  of nonlinear equations of
electrodynamics describing the motion  of  electrical  charge  in  the
field generated by the motion of the this charge.

We may find from here that wave equations for the  vector  and  scalar
potentials take the following form under Lorentz gauge
$(1/c)\partial\phi/\partial t+\nabla\cdot{\boldsymbol A}=0$:
\[
\Box{\boldsymbol A}-\frac{{\dot c}}{c^3}\frac{\partial {\boldsymbol A}}{\partial t}=
4\pi\rho\frac{{\boldsymbol v}}{c},\qquad
\displaystyle
\Box\phi-\frac{{\dot c}}{c^3}\frac{\partial\phi}{\partial t}=
4\pi\rho,
\]
where ${\dot c}=dc/dt={\boldsymbol v}\cdot{\dot{\boldsymbol v}}/c$. For the free field
with $\rho=0$  we have ${\dot c}=0$,  $v^2={\rm const}$,  ${\boldsymbol v}\cdot{\dot
{\boldsymbol v}}=0$, $\Box{\boldsymbol A}=0$, $\Box\phi=0$. When $c=c_0$ they coincide
with the equations from SR.

The condition $\nabla c=0$ imposes  certain  limits  on  the  possible
movement  of  a  particle.  Acting  by  the  operator  $\nabla$ on the
equation $\dot c=(e/mc_0){\boldsymbol v}\cdot{\boldsymbol E}$, we find
\begin{gather*}
\rho\left[({\boldsymbol v}\cdot\nabla){\boldsymbol E}+({\boldsymbol E}\cdot\nabla){\boldsymbol v}-
\frac{1}{c}{\boldsymbol v}\times \frac{\partial {\boldsymbol H}}{\partial t}\right]\\
\qquad{}+ \frac{c}{4\pi}{\boldsymbol E}\times \left(\Box{\boldsymbol H}-\frac{\dot c}{c^3}
\frac{\partial {\boldsymbol  H}}{\partial     t}\right)-\frac{mc_0}{e}{\dot
c}\nabla\rho+{\boldsymbol v}({\boldsymbol E}\cdot\nabla\rho)=0.
\end{gather*}
For a free particle (${\dot c}=0$,  $\rho=0$,  $e$=0)  we  have  ${\boldsymbol
E}\times \Box{\boldsymbol H}=0$, ${\dot {\boldsymbol v}}=0$.

As follows  from  the  equations   $\nabla\cdot{\boldsymbol   E}=4\pi\rho$, $
c\nabla \times {\boldsymbol H}-\partial_t{\boldsymbol E}
=4\pi\rho{\boldsymbol v}$ from the set
(\ref{kotel'nikov:eq24}),  in  the  present  theory  the  law  of  the
electrical charge conservation is valid:
\[
\frac{\partial\rho}{\partial t}+\nabla\cdot(\rho{\boldsymbol v})=0.
\]

Following \cite{kotel'nikov:landau&lifshitz},  we   find   the   well-known
relation:
\[
\frac{\partial}{\partial      t}\left(\frac{E^2+H^2}{8\pi}\right)=-c{\boldsymbol
j}\cdot{\boldsymbol E}-\nabla\cdot{\boldsymbol S}.
\]
Here $W=(E^2+H^2)/8\pi$  is  the  energy  density  of  electromagnetic
field, ${\boldsymbol    S}=(c/4\pi){\boldsymbol    E}\times
{\boldsymbol     H}=(c(0){\cal E}(t)/4\pi{\cal E}(0)){\boldsymbol E}\times
{\boldsymbol H}$ is the Poynting vector.

At last taking into account the expression for the velocity of light
\[
c(t)=c_0\sqrt{1+\frac{v^2(t)}{{c_0}^2}}=c(0)\left[1+\frac{e}{mc_0c(0)}
\int_{0}^{t}{\boldsymbol v}\cdot{\boldsymbol E}d\tau\right]=
c(0)\left[1+\frac{{\cal E}(t)-{\cal E}(0)}{{\cal E}(0)}\right],
\]
where $c(0)$ is the velocity  of  light  at  the  time  $t=0$,  ${\cal
E}(0)=mc_0c(0)$,  we  may  rewrite the sets (\ref{kotel'nikov:eq22}),
(\ref{kotel'nikov:eq24}) in the equivalent form:
\begin{gather}
m\frac{d{\boldsymbol v}}{dt}=
\frac{c(0)}{c_0}\left[1+\frac{ {\cal E}(t)-{\cal E}(0) }{{\cal E}(0) }
\right]e{\boldsymbol E}+\frac{e}{c_0}{\boldsymbol v}\times {\boldsymbol H},\nonumber \\
\frac{d{\cal E}}{dt}=e{\boldsymbol v}\cdot{\boldsymbol E}\to{\cal E}(t)-{\cal E}(0)=
e\int_{0}^{t}{\boldsymbol v}\cdot{\boldsymbol E}d\tau;\label{kotel'nikov:eq26}
\\
\left[1+\frac{ {\cal E}(t)-{\cal E}(0) }{{\cal E}(0)}\right]\nabla \times
{\boldsymbol E}+\frac{1}{c(0)}\frac{\partial{\boldsymbol H}}{\partial t}=0,    \qquad
\nabla\cdot{\boldsymbol E}=4\pi\rho,                                        \\
\left[1+\frac{{\cal E}(t)-{\cal E}(0)}{{\cal E}(0)}\right]
\nabla\times {\boldsymbol H}-\frac{1}{c(0)}\frac{\partial{\boldsymbol E}}{\partial t}=
4\pi\rho\frac{{\boldsymbol v}}{c(0)},\qquad
\nabla\cdot{\boldsymbol H}=0.\label{kotel'nikov:eq26a}
\end{gather}
The set (\ref{kotel'nikov:eq26}),  (\ref{kotel'nikov:eq26a})  coincide
with    the    set    of    equations   of   Maxwell   electrodynamics
\cite{kotel'nikov:landau&lifshitz} in the approximation $[{\cal E}(t)-{\cal
E}(0)]/{\cal E}(0)\ll 1$ with $c(0)=c_0$.

Further let   us   pay   attention   to   the   expression   $v=({\cal
E}^2-{m}^2{c_0}^4)^{1/2}/mc_0>c_0$.  It follows from here that in  the
framework of   the   present   work   a   free   particle   will  move
faster-than-light,  if the particle  energy  satisfies  the  condition
\cite{kotel'nikov:kotel'nikov}
\begin {equation}\label{kotel'nikov:eq27}
{\cal E}>\sqrt2m{c_0}^2=\sqrt2{\cal E}_0.
\end{equation}
The energy ${\cal E}=2^{1/2}{\cal E}_0$ is equal $\sim 723~{\rm  keV}$
for  the electron (${\cal E}_0\sim 511~{\rm keV}$) and $\sim 1327~{\rm
Mev}$ for the proton and neutron (${\cal E}_0\sim 938~{\rm MeV}$).  We
may conclude from here that in the present work the neutron physics of
nuclear reactors may be formulated in the approximation $v<c_0$ as  in
SR.  The  electrons with the energy ${\cal E}>723~{\rm keV}$ should be
faster-than-light particles (for example,  the velocity of the $1$~GeV
 electron  should  be $\sim 2000~c_0$).  The particle physics on
accelerators with the energy of protons  more  than  $1.33~{\rm  GeV}$
would be physics of faster-than-light motions,  if the present  theory
be realized in the nature.

\section{Application to physics}
Let us consider how  a  set  of  the  well-known  experiments  may  be
interpreted  in  the  framework  of the present theory.  As an example
chosen may be: the Michelson experiment, Fizeau experiment, aberration
of  light,  appearance of atmospheric $\mu$-mesons near the surface of
the Earth,  Doppler effect,  known tests to check independence of  the
speed  of  light from the velocity of light source,  decay of unstable
particles,  creation of new particles in nuclear  reactions,  possible
faster-than  light  motion  of  nuclear  reactions  products,  Compton
effect, photo-effect. Below we shall consider some of them.

{\bf Michelson experiment.}
Negative result of the Michelson experiment for  an  observer  with  a
terrestrial source  of  light  (the reference frame $K$,  the speed of
light $c_0$) may be explained by space isotropy.  Since the  speed  of
light $c_0$ is the same in all directions, a shift of the interference
pattern is absent with the interferometer's rotation.  Analogously, in
the  case  of an extraterrestrial source (the reference frame $K'$ $-$
the star moving inertialy  with  a  velocity  $v$  relatively  to  the
Earth),     the     velocity     of     light     from     the    star
$c=c_0(1+v^2/{c_0}^2)^{1/2}$ is the same for an observer on the Earth.
As  a  result,  the  interference  pattern  does  not  change with the
interferometer's rotation \cite{kotel'nikov:kotel'nikov}.

{\bf Aberration of light.}
By analogy with  SR  \cite{kotel'nikov:pauli,  kotel'nikov:landau&lifshitz}  for
one-half   of   the   aberration  angle  (in  arc  seconds)  we  have:
$\sin\alpha=V/c$,
$\alpha\sim(V/c_0)(c_0/c)\sim20.5(1+z_{\omega})/(1+z_{\lambda})$
\cite{kotel'nikov:kotel'nikov}.      Here       $z_{\omega}=(c-Vn_x)/c_0-1,
z_{\lambda}=(c-Vn_x)c/{c_0}^2-1$   are  the  redshift  parameters  for
frequency and wavelength  respectively.  It  follows  from  here,  for
quasar Q$1158+4635$ with $z_{\lambda}=4.73$ (Carswell and Hewett,  1990)
we obtain at $n_x=-1$,  that $z_{\omega}\sim 2.23$,  $c\sim  1.77c_0$,
$\alpha\sim  11.6$  instead  of being $z_{\omega}=z_{\lambda}$,  $c=c_0$,
$\alpha\sim 20.5$ in SR.

{\bf Appearance of atmospheric $\boldsymbol{\mu}$-mesons near the  surface  of  the
Earth.}
Since the  time  dilatation  is  absent  in  the  present  work,   the
appearance  of air $\mu$-mesons near the surface of the Earth could be
explained by faster-than-light motion of the mesons with  velocity  of
the order of $6\cdot10^6/2.2\cdot10^{-6}\sim3\cdot10^{10}$ m/sec,  or
$100c_0$.   This   corresponds   to   the    meson    energy    ${\cal
E}_{\mu}=m_{\mu}c_0c\sim10.6$~GeV,  where $m_{\mu}{c_0}^2\sim106$~MeV
is the rest energy of $\mu$-meson \cite{kotel'nikov:kotel'nikov}.

{\bf Tests to check independence  of  the  speed  of  light  from  the
velocity of light source.}
It is,  for example,  Bonch--Bruevich--Molchanov (1956)  experiment,  in
which  the  velocities  of  light  radiated by the eastern and western
equatorial edges of the solar disk  were  compared;  Sadeh  experiment
(1963) in which the velocities of $\gamma$-quanta, arising as a result
of  the  electron-positron  annihilation  in  flight,  were  compared;
Filippas--Fox  experiment  (1964),  where the effect of the velocity of
fast $\pi^0$-meson on the velocities  of  $\gamma$-quanta  from  decay
$\pi^0\to\gamma+\gamma$ were investigated.  In view of independence of
the velocity of light (\ref{kotel'nikov:eq4}) from  the  direction  of
emission  of light relatively to the vector of velocity ${\boldsymbol v}$ of a
light source (the solar disk, center-of-mass of electron and positron,
$\pi^0$-meson)  the  result  in  such  type  of  experiments should be
negative \cite{kotel'nikov:kotel'nikov}.

{\bf Decay  of  unstable  particles.}
Because of the equality of the rest energy of particles in SR  and  in
the present work,  the condition of spontaneous decay of the particles
into fragments is the same for the both theories.  In particular,  for
the  case  of decay of a particle with the mass $M$ into two fragments
with the masses $m_1$ and $m_2$,  the energy conservation law leads to
$M{c_0}^2={\cal E}_1+{\cal E}_2$,  where ${\cal E}_1$ and ${\cal E}_2$
are  the  energies   of   the   particles   produced.   Since   ${\cal
E}_1>m_1{c_0}^2$, ${\cal E}_2>m_2{c_0}^2$,  the decay is possible
if $M>m_1+m_2$  (as  in  SR~\cite{kotel'nikov:landau&lifshitz}).  From  the
conservation  law of energy-momentum $M{c_0}^2={\cal E}_1+{\cal E}_2$,
${\boldsymbol p}_1 + {\boldsymbol p}_2=0$,  and relations (\ref{kotel'nikov:eq20})  it
follows                           that                          ${\cal
E}_1={(M^2+{m_1}^2-{m_2}^2)c_0^2}/{2M}$,        ${\cal
E}_2={(M^2-{m_1}^2+{m_2}^2)c_0^2}/{2M}$     as     in    SR
\cite{kotel'nikov:landau&lifshitz}.  The difference consists in  predicting
the  fragment  velocities.  Using formula ${\cal E}=mc_0c$,  for these
velocities      in      the      present      work      we      obtain
$v_1=[((M^2+{m_1}^2-{m_2}^2)^2/4{m_1}^2M^2)-1]^{1/2}c_0$,
$v_2=[((M^2-{m_1}^2+{m_2}^2)^2/4{m_2}^2M^2)-1]^{1/2}c_0$.  One can see
that    when    ${(M^2+{m_1}^2-{m_2}^2)^2}/{4{m_1}^2M^2}\geq   2$,
${(M^2-{m_1}^2+{m_2}^2)^2}/{4{m_2}^2M^2}\geq 2$,  the velocities $v_1$
and   $v_2$   can  exceed  or  be  equal  the  speed  of  light  $c_0$
\cite{kotel'nikov:kotel'nikov}.

{\bf Creation  of new particles.}
Let us   consider   the   reaction   of  antiproton  creation  in  the
proton-proton collision in a laboratory reference frame $K$: $p^+({\rm
moving})+p^+({\rm  in~rest})=p^+ +p^+ +p^+ +p^-$,  where we denote the
energy of the moving proton as ${\cal E}_1$,  the  momentum  as  ${\boldsymbol
p}_1$,     and     the     proton    mass    as    $m_p$.    Following
\cite{kotel'nikov:landau&lifshitz} and using the relationship  between  the
momentum  and  energy (\ref{kotel'nikov:eq20}),  one can write $({\cal
E}_1+m_p{c_0}^2)^2-{c_0}^2{p_1}^2=16{m_p}^2{c_0}^4$.  In view  of  the
relationship   ${{\cal   E}_1}^2-{c_0}^2{p_1}^2={m_p}^2{c_0}^4,$   one
obtains $2\,{\cal  E}_1m_p{c_0}^2=14\,{m_p}^2{c_0}^4$.  From  here  we
find that  the  threshold  energy  of  the antiproton creation ${\cal
E}_1=7{m_p}{c_0}^2\sim   7$~GeV    is    the    same    as    in    SR
\cite{kotel'nikov:taylor&wheeler}.  The difference consists in the value of
velocities of particles.  In  particular,  according  to  the  formula
$p_1=m_pv_1=({{\cal E}_1}^2/{c_0}^2-{m_p}^2{c_0}^2)^{1/2}=(49-1)^{1/2}
m_p{c_0}$,  in the present work the velocity of proton possessing  the
energy         $7$~GeV        is        $v_1=(48)^{1/2}{c_0}=6.9{c_0}$
\cite{kotel'nikov:kotel'nikov} is distinct from SR.

{\bf Compton effect and photo-effect.}
Following \cite{kotel'nikov:blokhintsev49},     we     find     from    the
energy-momentum   the  conservation     law      (\ref{kotel'nikov:eq19})
$\hbar\omega=\hbar\omega'+m_e{c_0}^2 [(1+{v^2}/{{c_0}^2})^{1/2}-1]$;
$ {\hbar\omega}/{c_0}=({\hbar\omega'}/{c_0})\cos\theta+m_ev\cos\alpha$,
$0=({\hbar\omega'}/{c_0})\sin\theta-m_ev \sin\alpha$.  Here $\hbar\omega,
\  \hbar\omega'$  are  the  energies   of   incident   and   scattered
$\gamma$-quanta,  $\alpha$  and  $\theta$ are the angles of scattering
the electron and $\gamma$-quantum respectively,  $m_e$ is the electron
mass.   As   a   result   the   angular   distribution   of  scattered
$\gamma$-quanta                                                     is
$\omega'=\omega/[1+\hbar\omega(1-\cos\theta)/m_e{c_0}^2]$     as     in
\cite{kotel'nikov:blokhintsev49}.  But the  velocity  of  forward-scattered
electron     may     exceed     the     speed    of    light    $c_0$:
$v_e= c_0(\hbar\omega/m_e{c_0}^2)[1-m_e{c_0}^2/
(2\hbar\omega+m_e{c_0}^2)]>
c_0$ if $\hbar\omega>698$~keV,   which  differs  from  SR.  The
velocity of scattered $\gamma$-quantum does not depend  on  the  angle
$\theta$  and  is  determined  by mechanism of scattering (immediately
after interaction of $\gamma$-quantum with electron,  or in the act of
absorption-emission by scattered electron).

Analogously, in the case of photo-effect the velocity of photoelectron
is  equal  $v=c_0[((\hbar\omega+m_e{c_0}^2-U)/m_e{c_0}^2)^2-1]^{1/2}$,
where  $U$  is  the  energy  of  ionization.  If  the energy of photon
$\hbar\omega\ge (2^{1/2}-1)m_e{c_0}^2+U=211~{\rm keV}+ U$,  the velocity of
photoelectron is $v\ge c_0$.

\newpage

\section{Conclusion}
The $L_6\times \triangle_1$  invariant  theory  has been constructed,
where  $L_6$  is  the  Lorentz  group,  $\triangle_1$  is  the   scale
transformation  group  of  the  velocity  of  light $c'=\gamma c$.
In accordance with Blokhintsev papers \cite{kotel'nikov:blokhintsev46}
we may  assume  that the proposed theory may prove to be useful in the
field of particles physics,  when the elementary  particle  is  not  a
point but possesses some dimensions.
 Indeed,
the  elementary  particles  should  be  points  in the $L_6$ invariant
theory (SR) because of the finiteness of the speed of light $c_0$.  In
the  $L_6\times \triangle_1$ theory this requirement is not necessary
because of the absence of the limit to the velocity of light $c$.  The
postulation $c'=c$ leads to SR.

\LastPageEnding

\end{document}